\documentclass[longauth]{aa} 
\usepackage{graphicx}
\usepackage{txfonts}
\newcommand{\um}{$\mu$m}
\newcommand{\cmt}{cm$^{-3}$}
\newcommand{\cmd}{cm$^{-2}$}
\newcommand{\wat}{H$_2$O}
\newcommand{\h}{H$_2$}
\newcommand{\kms}{km\,s$^{-1}$}

\begin{document}
\title{Water cooling of shocks in protostellar outflows: 
}

   \subtitle{\textit{Herschel}-PACS map of L1157\thanks{\textit{Herschel} is an ESA space observatory with science instruments provided by European-led Principal Investigator consortia and with important partecipation from NASA}}

   \author{B. Nisini\inst{1} 
   \and M. Benedettini\inst{2} 
   \and C. Codella\inst{3} 
   \and T. Giannini\inst{1}
   \and R. Liseau\inst{4} 
\and D. Neufeld\inst{5} 
   \and M. Tafalla\inst{6} 
   \and E. F. van Dishoeck\inst{7,8} 
   \and R. Bachiller\inst{6} 
   \and A. Baudry\inst{9} 
   \and A.~O. Benz\inst{10} 
   \and E. Bergin\inst{11} 
   \and P. Bjerkeli\inst{4} 
   \and G. Blake\inst{12} 
   \and S. Bontemps\inst{9} 
   \and J. Braine\inst{9} 
   \and S. Bruderer\inst{10}
   \and P. Caselli\inst{13,3} 
   \and J. Cernicharo\inst{14}  
   \and F. Daniel\inst{14} 
   \and P. Encrenaz\inst{15} 
   \and A.M. di Giorgio\inst{2} 
   \and C. Dominik\inst{16,17} 
   \and S. Doty\inst{18} 
   \and M. Fich\inst{19} 
   \and A. Fuente\inst{6} 
   \and J.R. Goicoechea\inst{14} 
   \and Th. de Graauw\inst{20} 
   \and F. Helmich\inst{20} 
   \and G. Herczeg\inst{8} 
   \and  F. Herpin\inst{9} 
   \and M. Hogerheijde\inst{7} 
   \and  T. Jacq\inst{9} 
   \and D. Johnstone\inst{21,22}    
   \and J. J{\o}rgensen\inst{23} 
   \and  M. Kaufman\inst{24} 
   \and L. Kristensen\inst{7} 
   \and  B. Larsson\inst{25} 
   \and  D. Lis\inst{12} 
   \and  M. Marseille\inst{20} 
   \and C. McCoey\inst{19} 
   \and G. Melnick\inst{26} 
   \and M. Olberg\inst{4} 
   \and B. Parise\inst{25} 
   \and J. Pearson\inst{28} 
   \and R. Plume\inst{29} 
   \and  C. Risacher\inst{20} 
   \and J. Santiago\inst{6} 
   \and  P. Saraceno\inst{2} 
   \and R. Shipman\inst{20}
   \and T.A. van Kempen\inst{26} 
    \and  R. Visser\inst{7} 
   \and S. Viti\inst{30,2}     
   \and S. Wampfler\inst{10} 
   \and F. Wyrowski\inst{27}             
   \and F. van der Tak\inst{20,31} 
   \and U.A.~Y{\i}ld{\i}z\inst{7}
  \and B. Delforge\inst{32,17}  
  \and J. Desbat\inst{9,33} 
  \and W.A. Hatch\inst{29} 
\and I. P\'eron\inst{34,32,17} 
\and R. Schieder\inst{35} 
\and J.A. Stern\inst{29}
\and D. Teyssier\inst{36} 
\and N. Whyborn\inst{37} }
   \institute{
INAF - Osservatorio Astronomico di Roma, Via di Frascati 33, 00040 Monte Porzio Catone, Italy:
              \email{nisini@oa-roma.inaf.it} 
\and 
INAF - Istituto di Fisica dello Spazio Interplanetario, Area di Ricerca di Tor Vergata, via Fosso del Cavaliere 100, 00133 Roma, Italy
\and 
INAF – Osservatorio Astrofisico di Arcetri, Largo E. Fermi 5, 50125 Firenze, Italy
\and 
Department of Radio and Space Science, Chalmers University of Technology, Onsala Space Observatory, 439 92 Onsala, Sweden
\and 
Department of Physics and Astronomy, Johns Hopkins University, 3400 North Charles Street, Baltimore, MD 21218, USA
\and 
IGN Observatorio Astron\'{o}mico Nacional, Apartado 1143, 28800 Alcal\'{a} de Henares, Spain
\and 
Leiden Observatory, Leiden University, PO Box 9513, 2300 RA Leiden, The Netherlands
\and 
Max Planck Institut for Extraterestrische Physik, Garching, Germany
\and
Universit\'{e} de Bordeaux, Laboratoire d’Astrophysique de Bordeaux, France; CNRS/INSU, UMR 5804, Floirac, France
\and
Institute of Astronomy, ETH Zurich, 8093 Zurich, Switzerland
\and
Department of Astronomy, The University of Michigan, 500 Church Street, Ann Arbor, MI 48109-1042, USA
\and
California Institute of Technology, Division of Geological and Planetary Sciences, MS 150-21, Pasadena, CA 91125, USA
\and
School of Physics and Astronomy, University of Leeds, Leeds LS2 9JT
\and
Centro de Astrobiolog\'{\i}a. Departamento de Astrof\'{\i}sica. CSIC-INTA. Carretera de Ajalvir, Km 4, Torrej\'{o}n de Ardoz. 28850, Madrid, Spain.
\and
LERMA and UMR 8112 du CNRS, Observatoire de Paris, 61 Av. de l'Observatoire, 75014 Paris, France
\and
Astronomical Institute Anton Pannekoek, University of Amsterdam, Kruislaan 403, 1098 SJ Amsterdam, The Netherlands  
\and
Department of Astrophysics/IMAPP, Radboud University Nijmegen, P.O. Box 9010, 6500 GL Nijmegen, The Netherlands
\and
Department of Physics and Astronomy, Denison University, Granville, OH, 43023, USA
\and
University of Waterloo, Department of Physics and Astronomy, Waterloo, Ontario, Canada
\and
SRON Netherlands Institute for Space Research, PO Box 800, 9700 AV, Groningen, The Netherlands
\and
National Research Council Canada, Herzberg Institute of Astrophysics, 5071 West Saanich Road, Victoria, BC V9E 2E7, Canada
\and
Department of Physics and Astronomy, University of Victoria, Victoria, BC V8P 1A1, Canada
\and
Centre for Star and Planet Formation, Natural History Museum of Denmark, University of Copenhagen,
{\O}ster Voldgade 5-7, DK-1350 Copenhagen, Denmark
\and
Department of Physics and Astronomy, San Jose State University, One Washington Square, San Jose, CA 95192, USA
\and
Department of Astronomy, Stockholm University, AlbaNova, 106 91 Stockholm, Sweden
\and
Harvard-Smithsonian Center for Astrophysics, 60 Garden Street, MS 42, Cambridge, MA 02138, USA
\and
Max-Planck-Institut f\"{u}r Radioastronomie, Auf dem H\"{u}gel 69, 53121 Bonn, Germany
\and
Jet Propulsion Laboratory, California Institute of Technology, Pasadena, CA 91109, USA
\and
Department of Physics and Astronomy, University of Calgary, Calgary, T2N 1N4, AB, Canada
\and
Department of Physics and Astronomy, University College London, Gower Street, London WC1E6BT
\and
Kapteyn Astronomical Institute, University of Groningen, PO Box 800, 9700 AV, Groningen, The Netherlands
\and
Institute Laboratoire d'Etudes du Rayonnement et de la Matière en Astrophysique, UMR 8112  CNRS/INSU, OP, ENS, UPMC, UCP, Paris, France 
\and
CNRS/INSU, UMR 5804, B.P. 89, 33271 Floirac cedex, France 
\and
Institute Institut de Radioastronomie Millimetrique, IRAM, 300 rue de la Piscine, F-38406 St Martin d'Heres 
\and
KOSMA, I. Physik. Institut, Universit\"at zu K\"oln, Z\"ulpicher Str. 77, D 50937 K\"oln
\and
European Space Astronomy Centre, ESA, P.O. Box 78, E-28691 Villanueva de la Cañada, Madrid
\and
ALMA
}

 
  \abstract
  {The far-IR/sub-mm spectral mapping facility provided by the \textit{Herschel}-PACS and HIFI instruments has made it possible 
  to obtain, for the first time, images of \wat\ emission with a spatial 
  resolution comparable to ground based mm/sub-mm observations.}
%
   {In the framework of the Water in Star-forming regions with \textit{Herschel} (WISH) key program, 
   maps in water lines of several outflows from young stars are being obtained, to study the water production in shocks and its role in the outflow cooling. This paper reports the first
   results of this program, presenting a PACS map of the o-\wat\ 179~\um\, transition
   obtained toward the young outflow L1157.}
   {The 179~\um\, map is compared with those of other important shock tracers, and with previous 
   single-pointing ISO, SWAS, and Odin water observations of the same source that allow us to constrain the
   \wat\ abundance and total cooling.
   }
   {Strong \wat\, peaks are localized on both shocked emission knots and the central source position. The 
   \wat\, 179~\um\, emission is spatially correlated with emission from \h\ rotational lines, excited
   in shocks leading to a significant enhancement of the water abundance. 
   Water emission peaks along the 
   outflow also correlate with peaks of other shock-produced molecular species, such as SiO and NH$_3$.
   A strong \wat\,peak is also observed at the location of the proto-star, where none of the other molecules have significant emission.
   The absolute 179~\um\, intensity and its intensity ratio to the \wat\ 557~GHz line previously observed with Odin/SWAS 
   indicate that the water emission originates in warm compact clumps, spatially unresolved by PACS, 
   having a \wat\, abundance of the order of 10$^{-4}$. This testifies that the clumps have been heated for a time
   long enough to allow the conversion of almost all the available gas-phase oxygen into water.   
   The total \wat\ cooling is $\sim$\,10$^{-1}$\, L$_\odot$, about 40\%
   of the cooling due to \h\ and 23\% of the total energy released in shocks along the L1157 outflow.
   }
   {}


   \maketitle
%

\section{Introduction}
Among the main coolants in molecular shocks, water is the tracer most sensitive to physical variations and the temporal evolution of 
protostellar outflows, thus representing a very powerful probe of their shock conditions and thermal history (e.g., Bergin et al. 1998).
Water emission and excitation in shocks were studied extensively for the first time with the \textit{Infrared Space Observatory} (ISO), the first space facility with spectroscopic capabilities in the mid- and far-IR. 
ISO surveyed the water emission in a large sample of outflows from young stellar objects (YSOs), providing  a global statistical picture of the importance of water in the outflow cooling and of variations in its abundance with shock 
properties and ages (see e.g., Nisini 2003, van Dishoeck 2004). 
Following ISO, the SWAS and Odin facilities made it possible to observe the ortho-\wat fundamental line
at 557~GHz, providing important constraints on the water abundance and kinematics in the cold outflow  gas components (e.g., Franklin et al. 2008, Bjerkeli et al. 2009).
All these facilities, however, had poor spatial resolution (i.e. greater than 80$\arcsec$), which did not allow one to locate the origin of the water emission nor study variations in abundances and excitation within individual flows.

In this framework, a sample of YSO outflows will be surveyed in different water lines by the PACS and HIFI instruments onboard the \textit{Herschel} satellite, as part of the key program WISH (Water In Star-forming-regions with \textit{Herschel}\footnote{http://www.strw.leidenuniv.nl/WISH/.}).
This paper presents the first results obtained 
from this survey, consisting of a PACS map of the H$_{2}$O 2$_{12}$-1$_{01}$ 179~\um\ line covering the outflow of the protostar L1157-mm, obtained during the \textit{Herschel} Science Demonstration Phase.
The 179~\um\ line is the transition connecting the lower two back-bone levels of ortho-\wat. It is 
therefore one of the brightest water lines expected in collisionally excited conditions, thus
representing an ideal tracer of the water distribution in shocked regions. L1157 is a well known outflow driven by a low mass class 0 object  (L1157-mm, $ L_{bol} \sim
8.3$ L$_\odot$, D=440 pc, Froebrich 2005). It is considered to be the prototype of chemically active flows, given the large number of different species 
detected in its shocked regions (e.g., Bachiller \& Perez-Gutierrez 1997). This paper is presenting the first of several observations planned
for this source by the WISH team.
%
%


\section{Observations}

Observations were performed on 26 October 2009 with the PACS instrument (Poglitsch et al. 2010) 
onboard the
\textit{Herschel} Space Observatory (Pilbratt et al. 2010) in line spectroscopic mode, with the grating centred 
on the \wat\ 2$_{12}$-1$_{01}$ line at 179.527~\um. The L1157 outflow region (of about $6\arcmin \times 2\arcmin$) was covered by 3 individual PACS raster maps, arranged along the outflow
axis. Each map consists of 3$\times$3 PACS frames acquired in steps of 40$\arcsec$. 
The instrument is a $5 \times 5$ pixel array providing a spatial sampling of 9.4$\arcsec$/pixel, 
while the spectral resolution at 179~\um\ is
$R\sim$1500 (i.e., $\sim$210 km\,s$^{-1}$). 
The data were reduced with HIPE 2.0. 
Additional IDL routines were developed to construct a final integrated and continuum-subtracted line map.
Flux calibrations used calibration files obtained by ground tests that remain very uncertain at the time of paper writing, especially for extended sources. To evaluate the flux uncertainty, we compared with 
the three measurements performed by the ISO satellite along the outflow (Giannini et al. 2001). To do that,
we performed aperture photometry of the line emission in the PACS map within the 80$\arcsec$ ISO circular beam. 
The ratio of PACS to ISO fluxes ranges between 1.1 and 1.8 at the three positions: we adopt
this as the uncertainty in our quantitative analysis. The typical rms noise across the map is of
the order of 2\,10$^{-6}$ erg\,s$^{-1}$\,cm$^{-2}$\,sr$^{-1}$. 
  
 \begin{figure*}
\begin{center}
\includegraphics[angle=-90,width=\textwidth]{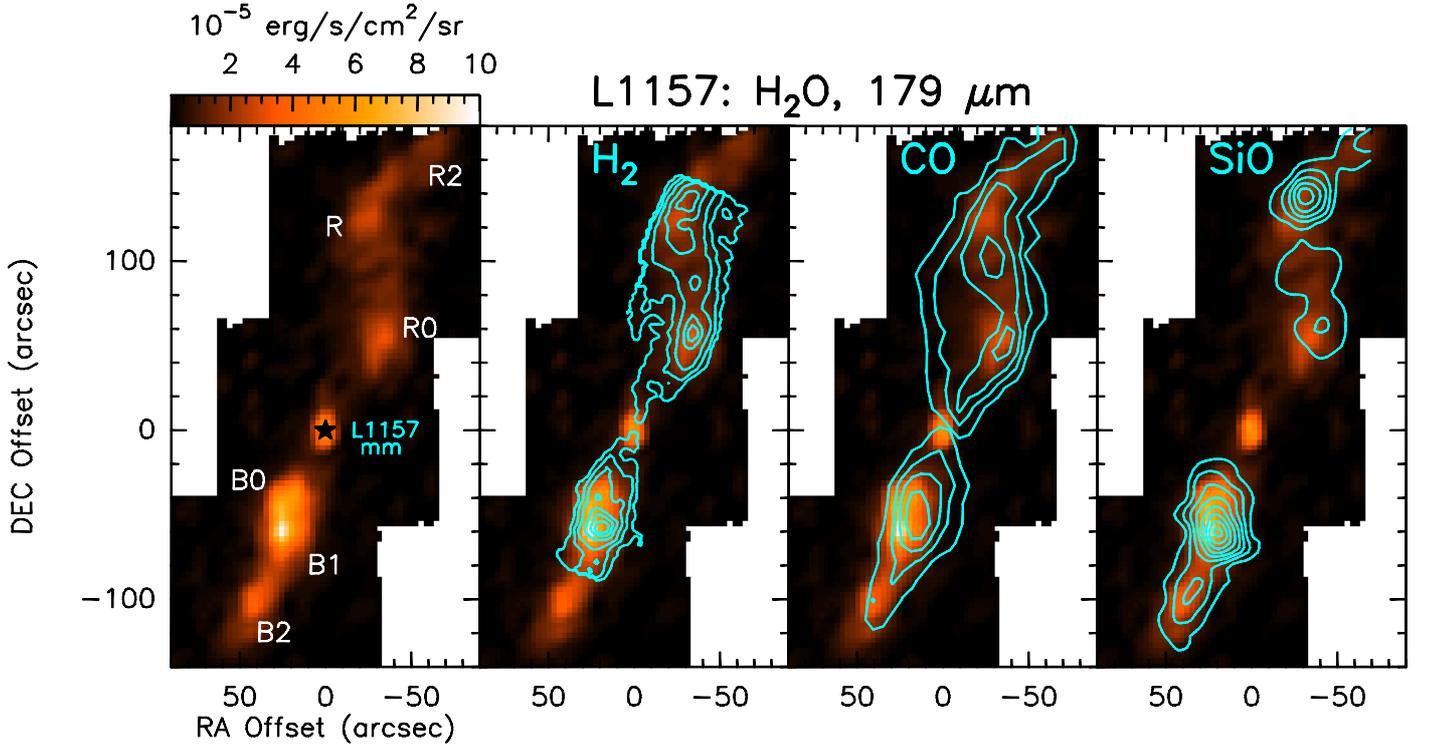}
\caption{Continuum subtracted PACS map of the integrated H$_2$O 179~\um\, emission along the L1157 outflow. 
Offsets are with respect to the L1157-mm source, at coordinates $\alpha(2000)$ = 20:39:06.2, $\delta(2000)$ =
$+$68:02:16. The different emission peaks are labelled following the nomenclature adopted by Bachiller et al. (2001)
   for individual CO peaks. The same map is shown in the other panels with overlays of other tracers, namely H$_2$ 0--0 S(1) at 17\um\, (Neufeld et al. 2009), CO 2-1, and SiO 3-2 (Bachiller et al. 2001).
   The spatial resolution of these images are $\sim$11\arcsec, for \h\, and CO, and 18\arcsec\, for SiO.
   Note that the \h\ observed region does not cover the B2 and R2 shocked peaks. }
\end{center}
\end{figure*}
%
\section{Results and comparison with other tracers}

Figure 1 presents the PACS map of the 179~\um\ line emission. In the same figure,
the \wat\, map is overlaid with contours of the emission from 
the \h\ 0--0 S(1) (Neufeld et al. 2009), CO 2-1 and SiO 3-2 (Bachiller et al. 2001) transitions.
The water map exhibits several emission 
peaks corresponding to the positions of previously-known shocked knots, 
labelled as B0-B1-B2 for the south east blue-shifted lobe, and R0-R-R2 for the north west red-shifted
lobe, following the nomenclature of Bachiller et al. (2001). These emission knots represent
 the actual working surfaces of a precessing and pulsed jet and are thus associated with the
present location of the active shock regions. With respect to CO, \wat\ emission appears more
localized, having a less prominent diffuse component. 
About 60\% of the total 179~\um\, flux is found within 30$\arcsec$ apertures centered on the knots. 
This could be partly related to the line excitation: the 179~\um\ line excitation temperature is $\sim$80 K above the o-H$_2$O ground state (compared to the $17$ K for CO 2-1),
and the critical density of its upper level is above 10$^8$ \cmt\, for $T \la$ 500 K. It may 
however also be a consequence
of the specific conditions needed to ensure a significant production of water. The \wat\ abundance
is indeed significantly higher only in shocks strong enough to release the water ice 
located on grain mantles by sputtering and grain-grain collisions or to activate the gas-phase reactions that convert  
the gas-phase oxygen into water. Both these  processes become efficient at shock velocities $v_s \ga$ 15 \kms\,
(Caselli et al. 1997, Jim{\'e}nez-Serra et al. 2008, Kaufman \& Neufeld 1996).
In this respect, we note that the \wat\ emission peaks correspond 
rather closely to both the position and the relative intensity of the \h\ rotational emission
(with the \wat~179~\um/\h~17~\um\, ratio in the range $\sim$(2--3)$\times 10^{-2}$ for all the \h\, peaks).
Peaks of low-$J$ \h\ pure rotational lines are associated with warm gas 
(with $T \sim$ 300-500 K) excited in low velocity non-dissociative shocks that are tracers 
of regions in which a high \wat\ abundance is expected.
Other molecules are known to have strongly enhanced abundances in shocks. One of the most well studied of these molecules
is SiO, for which Fig. 1 shows that, like water, its emission is very localized around the shocked knots. 
A similar behavior is found for other molecules, such as NH$_{3}$ and CH$_{3}$OH (Bachiller et al. 2001, Tafalla \& Bachiller 1995).

The strongest water peak is located at the position of the B1 knot, which is known to be
the most chemically active of the L1157 spots (e.g. Bachiller \& Perez Guitierrez 1997, Benedettini et al. 2007,
Codella et al. 2010). This knot at near-IR wavelengths appears 
as a bow shock with intense \h\ 2.12\um\ emission (Davis \& Eisl\"offel 1995) and has a significant \h\ column density
enhancement (Nisini et al. 2010). Although the spatial resolution of the present observations prevents us from 
completely resolving the bow shock structure, the observed morphology at the B1/B0 positions suggests that 
water emission is mainly localized at the bow apex and eastern wing. A similar morphology 
has been observed for molecules such as SiO, NH$_3$, and CS (Benedettini et al. 2007, Tafalla \& Bachiller 1997), while  
other shock produced molecules, such as CH$ _{3} $OH, noticeably have emission
localized on the bow western wing (e.g. Codella et al. 2009). This behavior probably relates to an asymmetry in the
excitation conditions along the bow structure, most likely induced by the jet precession or the 
propagation of shocks in an inhomogeneous medium.

Strong, spatially unresolved, water emission is also detected on-source.  This localized emission can originate
in different components, including shocks impacting on a dense medium at the jet base, the infalling protostellar envelope, or
emission from a UV-heated outflow cavity, as discussed in van Kempen et al. (2010) for the HH46-IRS case.
The precise origin of this emission will be investigated by dedicated \textit{Herschel} observations, but we note here the interesting  evidence that no other molecule exhibits significant emission at the central position. In particular, the non-detection of strong emission from molecules such as CH$_3$OH  indicates that grain ice mantle evaporation in the protostellar envelope is unlikely to be the origin of the on-source \wat\,  
emission, since the two molecules should desorb at similar temperatures. The non-detection of the \h\ 0--0 S(1) line at the central position is also remarkable. 
This may be caused by the heavy extinction close to the central source.
Assuming an intrinsic H$_2$O\,179\,\um/\h\,17\,\um\, ratio in the range of that observed along the outflow, we estimate that $A_v$ on-source should be $\ga$ 150 mag to be able to explain the \h\, line non-detection.
Alternatively, C-type shocks with very high pre-shock densities ($\ge$10$^6$\cmt) and velocities between 20 and 40 \kms\, are expected to have a large \wat/\h\, cooling ratio (Kaufman \& Neufeld 1996).
 
\section{Water abundance and total cooling}

To constrain the range of water column densities that could produce the observed 179~\um\ emission, we consider the SWAS and Odin observations of the \wat\ 1$_{10}$-1$_{01}$ 557~GHz (538\um)  line observed in this outflow (Franklin et al. 2008, Bjerkeli et al. 2009). Given the large size of the apertures of these two instruments relative to the PACS spatial resolution, we evaluate here only properties averaged over large outflow regions. 
In particular, we consider the Odin observations acquired towards the blue (B) and red (R) outflow lobes 
at offsets (+29$ \arcsec $,-52$ \arcsec $) and  (-21$ \arcsec $,+121$ \arcsec $) (Bjerkeli et al. 2009). The 179\,\um/557\,GHz 
 intensity ratios are obtained by diluting the PACS observations to the 126$\arcsec$ Odin resolution.
The same procedure was adopted for the SWAS observation that encompasses almost the entire L1157 PACS mapped region with its 3.5$^{\prime} \times$ 5.0$^{\prime}$ elliptical aperture.
\begin{figure}
\begin{center}
\includegraphics[angle=0,width=9cm]{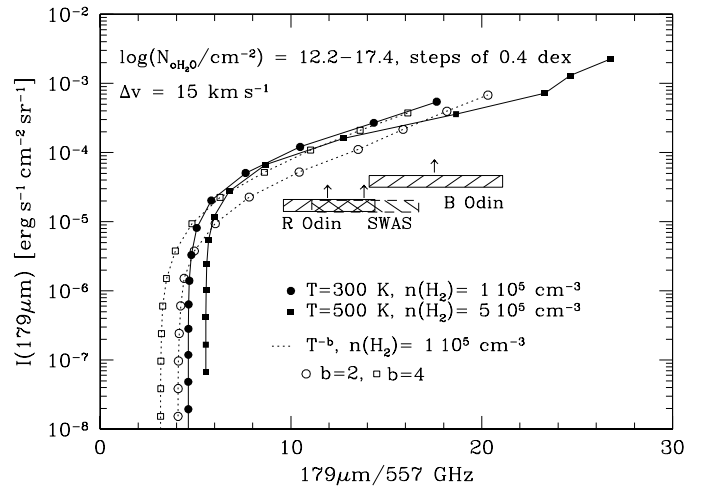}
\caption{LVG theoretical predictions of the 179~\um\ line brightness versus the 
179\,\um/557\,GHz line ratio, compared with observed values. See text for the details. }
\end{center}
\end{figure}
Figure 2 presents large velocity gradient (LVG) predictions, assuming a slab geometry, of the 179~\um\ line brightness versus the 
179\,\um/557\,GHz line ratio, compared to the observations combined above. 
The absolute brightnesses are those averaged within 
an area enclosing 90\% of the total PACS emission inside each considered Odin/SWAS aperture.
These emitting areas are 5.9$\times 10^{-8}$, 8.0$\times 10^{-8}$, and 2.7$\times 10^{-7}$ sr for the R, B, and the SWAS
apertures, respectively. The line intensity derived in this way was considered to be a lower limit to the 
true 179~\um\ brightness if the PACS emission originates in a clumpy medium, of which the clump size is
smaller than the \textit{Herschel} diffraction limit at 179~\um.

In the figure, observations are indicated as boxes that take into consideration the uncertainty of a factor 
of about 1.5 in the 179~\um\ flux, estimated by comparing with the ISO observations (Sect. 2).
Theoretical curves were derived as a function of the o-\wat\, column density, using the RADEX 
code (Van der Tak et al. 2007)
assuming temperature and density conditions measured from the \h\, \textit{Spitzer} observations
or ground-based millimeter observations (Nisini et al. 2010, Nisini et al. 2007, Mikami et al. 1992).
The temperature is between 300 and 500 K and the density is in the range 1--5$\times10^{5}$\cmt, the blue lobe
being on average colder and denser than the red lobe.
Part of the 557~GHz emission can arise from a gas colder than these assumed values, given the lower 
excitation temperature of this line with respect to the 179~\um\, line.
To evaluate the effect of different temperature components along the line of sight on 
the ratio of the two considered transitions,  
Fig.2  also plots the theoretical predictions assuming a temperature stratification where the 
column density in each layer at a given $T$ 
varies as $T^{-b}$ (Neufeld \& Yuan 2008). A minimum and maximum temperature of 100 K and 4000 K,
respectively are assumed, and $b$ values between 2 and 4, i.e., the range of values that consistently fit
the \h\ rotational lines (Neufeld et al. 2009). These curves give the same range of predicted values as the single $T$ curves, indicating that contributions from high-temperature gas do not
significantly affect the considered transitions.

Several general conclusions can be drawn from the inspection of Fig. 2. 
Firstly, the data are consistent with model predictions only if we assume 
that the real emitting areas are smaller than those estimated from the PACS map.
In particular, agreement with the theoretical curves is found for covering factors ($f_c$) 
$\sim$ 0.1-0.2, which suggests that the emission is concentrated on some unresolved emission knots
that together do not fill an area larger than a few tens of arcsec. This is not unexpected, since 
interferometric mm observations illustrate the extreme clumpiness of the shocked gas,  
individual knots being of sizes of a few arcsec each (e.g., Benedettini et al. 2007, Lefloch et al.
2010). We note that the typical length scale for planar C-type shocks at the considered 
densities is of the order of 10$^{16}$ cm,
i.e., about 1/10 of the PACS spatial resolution at D=440 pc.
The observed 179\,\um/557\,GHz ratios, ranging between 10 and 20, are consistent with 
$N$(\wat) $\sim$ 2--9$\times 10^{16}$ \cmd\, (assuming a $\Delta$v = 15 km\,s$ ^{-1}$ from the 557~GHz line width).
The \h\, column densities, averaged within the PACS emitting areas, were measured from
the \h\ mid-IR rotational lines and results in $\sim$ 5$\times 10^{19}$\cmd\, in
both regions covered by the B and R observations. 
The water abundance in the unresolved clumps is therefore estimated to be
$\sim N$(\wat)/$N$(\h)$\times f_c \sim 0.6-3\,10^{-4}$ (with a \wat\ o/p ratio of 3).
Table 1 reports in more detail the range of values derived in each considered aperture.
The total mass of the shocked gas involved in the 179~\um\, emission is of
the order of 5\,10$^{-3}$ M$_\odot$, which is only a small fraction ($\sim$ 1/100) of the total
mass of the outflow estimated from CO observations (e.g. Bachiller et al. 2001). Lefloch et al. (2010) show that \wat\, components with different velocities
are discernible in the 557~GHz data acquired by HIFI in a 40\arcsec beam centred on the L1157-B1 knot. 
They separately analyse the different velocity components, confirming
that small filling factors are required to explain their observations and finding that 
the component of higher velocity is the one exhibiting the water abundance of the order
of $10^{-4}$. Lower \wat\, abundance values, between 10$^{-6}$ and 10$^{-5}$, were estimated using only the SWAS and Odin 557~GHz emission, assuming that the 557~GHz emission originates in the same cool gas
traced by the low-$J$ CO emission, thus a gas with a larger covering factor and lower temperature than considered here (Neufeld et al. 2000, Franklin et al. 2008, Bjerkeli et al. 2009).
Combining ISO-179~\um\ emission and SWAS observations, Benedettini et al. (2002)
derived a water abundance for the warm shocked gas of $\sim$ 5\,10$^{-5}$, thus in the lower range of values estimated in the present analysis. However, the ISO observations
did not cover the entire L1157 outflow 179~\um\ emission,
and the inferred ISO\,179\,\um/SWAS\,557\,GHz ratio was underestimated by about a factor of 2.
\begin{table}{}
\caption{Estimated water abundances}
\begin{tabular}{lcccccc}
\hline
 &  $R^a$ &$T^b$ & $n$(H$_2)^b$ & $N$(oH$_2$O) & X(H$_2$O)$^b$ & $A^c$ 
\\ 
 & & K& \cmt & 10$^{16}$\cmd & 10$^{-4}$& sr \\
\hline
B & 14-20 &300 & 3\,10$^5$ &  3-9 & 0.6-3 & $\sim$ 5.2\,10$^{-9}$ \\
R & 10-14 &500 & 1\,10$^5$ &  2-4 & 0.8-2 & $\sim$ 9.4\,10$^{-9}$\\
\hline	    

\end{tabular}
$^a$ 179\,\um/557\,GHz ratio within the Odin aperture.  \\
$^b$ See text for references and assumptions on $T, n$ and $N$(\h).\\
$^c$ Effective emission area that reconciles the observed and predicted 179~\um\ line intensity within
the Odin aperture.
\end{table}
Given the considered conditions, the 179~\um\ line contributes to about 30-40\% of the water 
emission in the outflow: the total estimated \wat\ luminosity is $\sim$ 8-9$\times 10^{-2}$L$_\odot$, which is  
 about 40\% of the total \h\ shock luminosity (0.2 L$_\odot$, Nisini et al. 2010) and about 
 23\% of the total shock cooling in the L1157 outflow, if we also consider the contributions
 given by CO and [\ion{O}{i}] derived from ISO observations by Giannini et al. (2001).
The high water abundance estimated in the present analysis is consistent with predictions of
non-dissociative shock models, in which water is mainly produced
by endothermic reactions, activated at $T \ga$ 300 K, where all the available gas-phase oxygen 
is converted into \wat,  or by the sputtering of icy grain mantles behind the shock. According to Bergin et al. (1998), the time needed to complete this process is of the order of
10$^3$ yr, for $T$ = 400 K. This is comparable to the shock timescales estimated from \h\ observations of 
individual emission knots of the L1157 outflow (Nisini et al. 2010), thus supporting the idea that
the water in this outflow has had time to reach its maximum allowed abundance.

\section{Conclusions}
We have presented a PACS spectral map of the \wat\ 179~\um\ transition obtained toward the 
L1157 protostellar outflow. 
Strong water emission peaks have been found at the location of previously-known shocked spots
and correlate well with \h\ mid-IR rotational lines, as well as other important 
shock tracers, such as SiO and NH$_3$. The absolute 179~\um\, intensity and the intensity ratios with respect to the previously-observed 557~GHz line, indicate that the water emission originates in warm compact clumps, 
spatially unresolved by PACS, that have a \wat\, abundance of the order of 10$^{-4}$.
The total \wat\ cooling has been estimated to be of the order of 8-9\,10$^{-2}$L$_\odot$, representing about 40\% of the cooling due to \h\ and 23\% of the total energy released in shocks along the L1157 outflow.

Additional \textit{Herschel} PACS/HIFI observations of the L1157 outflow are planned  by the WISH program. 
These will enable us to investigate variations in the  water abundance within the outflow and 
correlate these with kinematical information.

\begin{acknowledgements}
 This program is made possible thanks to
the HIFI guaranteed time and the PACS instrument builders.
\end{acknowledgements}

\end{document}